\documentclass[12pt,a4paper]{article}
\usepackage{graphicx}

\usepackage{authblk}
\usepackage{fullpage,amsfonts,lineno,setspace}

\usepackage{hyperref}

\begin{document}

\title{The Fractured Nature of British Politics}

\author{Carlos Molinero\thanks{Corresponding
author\\ \href{mailto:c.molinero@ucl.ac.uk}{c.molinero@ucl.ac.uk}; \href{mailto:e.arcaute@ucl.ac.uk}{e.arcaute@ucl.ac.uk}; \href{mailto:duncan.a.smith@ucl.ac.uk}{duncan.a.smith@ucl.ac.uk}; \href{mailto:m.batty@ucl.ac.uk}{m.batty@ucl.ac.uk}}  }
\author{Elsa Arcaute }
\author{Duncan Smith }
\author{Michael Batty}
\affil{\small{Centre for Advanced Spatial Analysis (CASA)\\ University College London (UCL)}}

\date{\small{\today}}
\maketitle

\begin{abstract}
   The outcome of the British General Election to be held in just over one week's time (May 7th, 2015) is widely regarded as the most difficult in living memory to predict. Current polls suggest that the two main parties (Conservative and Labour) are neck and neck but that there will be a landslide to the Scottish Nationalist Party with that party taking most of the constituencies in Scotland (some 50 out of 59 on the most recent forecast for Sunday April 26th). The Liberal Democrats are forecast to loose more than half their seats (56 to 24) and the fringe parties of whom the UK Independence Party is the biggest are simply unknown quantities. Much of this volatility relates to long-standing and deeply rooted cultural and nationalist attitudes that relate to geographical fault lines that have been present for 500 years or more but occasionally reveal themselves, at times like this. In this paper our purpose is to raise the notion that these fault lines are critical to thinking about regionalism, nationalism and the hierarchy of cities in Great Britain (excluding Northern Ireland). We use a percolation method \cite{Arcaute2015} to reveal them that treats Britain as a giant cluster of related places each defined from the intersections of the road network at a very fine spatial scale (down to 50 metre resolution). We break this giant cluster into a detailed hierarchy of sub-clusters by successively reducing a distance threshold starting at 5 kilometres which first breaks off some of the Scottish Islands and then reveals the very distinct nations and regions that make up Britain, all the way down to the definition of the largest cities that appear when the threshold reaches 300 metres. We use these percolation clusters to apportion the 2010 voting pattern to a new hierarchy of constituencies based on these clusters, and this gives us a picture of how Britain might vote on purely geographical lines. We then examine this voting pattern which provides us with some sense of how important the new configuration of political parties might be to the election next week.
\end{abstract}


\section{Introduction}
Long-standing cultural differences between the countries that comprise the United Kingdom are part of the folklore of British politics but in the last twenty-five years they have reasserted themselves with a vengeance. The devolution of central powers from the government in Westminster to Scotland and Wales were first initiated a generation ago while Northern Ireland has had periods from the 1970s when its traditional parliament, now assembly, has been entirely suspended. However these divisions have not really asserted themselves in terms of national voting until the last decade but it now looks as though Britain in its forthcoming general election on 7th May 2015 will finally divide itself along very deep and historically significant lines.

In recent times, since the 1960s, the traditional two party system that had dominated politics since the early 20th century began to slowly fracture with the Liberals (now Liberal Democrats) reasserting themselves in south west England while also making inroads into somewhat conservative but relatively high income city suburbs and country towns.  The Scottish Nationalist Party (SNP) began to gain more seats after devolution and recent local and European elections have led to a massive increase in their support that is widely seen as a sea change in the quest for Scottish independence. The SNP are forecast to more or less wipe out the traditional Labour Party in Scotland in the forthcoming general election. The Welsh nationalists Plaid Cymru have a much smaller base in Wales although it is entirely possible that they will gain seats in May while in Northern Ireland the traditional focus on conservatism and nationalism has led to a split with the Conservative Party in England to which it is traditionally allied. Northern Ireland politics have become much more inward looking.

The two parties that have dominated national politics until quite recently the Conservatives (the Tories) and the Labour party have also changed. The Blair and Brown governments, from 1997 to 2010, introduced the philosophy that was called New Labour, with the party much more geared to contemporary business ethics and deregulation. There has been a substantial reaction against this with the party reverting to a somewhat more traditional stance but like the present coalition of Conservatives and Liberal Democrats espousing a pro-austerity stance in the wake of the Great Recession. The Liberals, of course, joined with the Conservatives in 2010 in a coalition, the first long lasting one for well over 70 years, and this has softened and blurred traditional thinking amongst the Conservatives and perhaps hardened and confused the philosophy of the Liberal Democrats. Add to this the emergence of the anti immigration and anti European Union party, UKIP (the United Kingdom Independence Party) and the picture at first sight appears more confused that at any time since the Labour Party began to erode support for the Liberals in the late 19th century.

Or is it? In our work on defining cities and regions within Britain (the UK less Northern Ireland), we are defining similar places by their connectivity to one another. Essentially we begin by treating Britain and all its places as a giant connected cluster that we define from the detailed road network that links places together, these places being defined at their most atomic level from the nodes where street segments intersect. The number of nodes of the  graph that defines this giant cluster is of the order of $\Vert V\Vert\simeq 3.3\cdot 10^6$ and the number of segments $\Vert E\Vert\simeq 4\cdot 10^6$ and when simplified as symmetric give an average degree for each node of $\langle k \rangle=2.34$. To decompose this network into clusters with different degrees of connectivity, we begin by specifying a range based on the maximum segment length in the network, gradually relaxing this threshold, thus producing a hierarchy of clusters which is a unique decomposition of the British geographical space. Note that we use street intersections as the basis for our definition of any spatial unit so a parliamentary constituency, an area for which a politician is elected, is regarded as all the nodes and segments that fall uniquely into the physical area defining that space.

The percolation method developed in \cite{Arcaute2015}  starts with the entire cluster, setting the starting threshold at $d = 5000$m and the hierarchy of clusters emerges as we successively reduce this value. As we might expect from a casual knowledge of Britain, the more remote periphery will disconnect first but we are not able to anticipate the actual partitioning. When the threshold reaches 1.4 km, Scotland breaks off completely from the rest of England and Wales. The break is very geographically distinct, with the central lowlands dividing the country entirely from England excluding a few of the English border counties in the south. When the threshold falls to 900m, the Industrial North and West and Wales separates off from the South East and then, when it falls another 100 metres, the South West and South Wales become distinct from this division. The big cities then fall out of this when the threshold falls to 300 metres. The hierarchy is so clear that it is hard not to conclude without knowing anything else about these regions, that they are culturally and economically quite distinct. In the light of the debate about Scottish independence and recent European election results, the geographical correlations with the predominant voting patterns are surprising and very clear. The influence of
geographical boundaries on voting dynamics has already been studied
in a few papers in the literature \cite{Johnston2006,Johnston2009,Perez2015} but the notion
of percolation and the way that it divides the geographical space vastly
improves our understanding on the matter and allows us to quantify those geographical units in an univocal manner.
Although there are many explorations of voting behaviours using physical concepts particularly in opinion dynamics and voting structure \cite{Torok2013}, and many that have used the concept of percolation applied to consensus decision making \cite{Stauffer2002,Shao2009,Balankin2011} and even studied the spread of opinions through networks \cite{Travieso2006,Lambiotte2007}, none, as far as we are aware, are using a geographical percolation to explain voting patterns.

\begin{figure}[h]
\center{\includegraphics[width=1\linewidth]{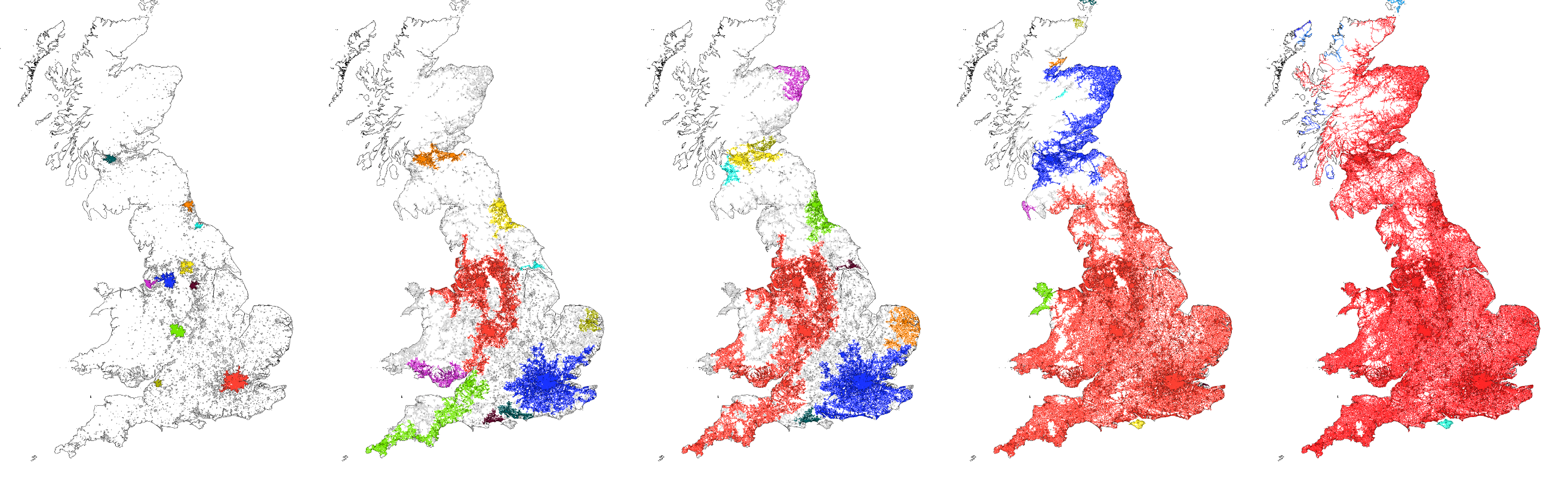}}
\caption{Network percolations at the main thresholds, respectively, 300m, 800m, 900m, 1400m and 5000m . To view the animated sequence of percolation distances please refer to: \url{http://www.mechanicity.info/percolation-clustering-of-the-uk-road-network/}.
}\label{percolations}
\end{figure}

To get an immediate idea of this kind of clustering, we refer you to our \href{http://www.mechanicity.info/percolation-clustering-of-the-uk-road-network/}{percolation movie} where we successively partition the space by increasing the distance threshold systematically. Figure~\ref{percolations}  shows the partition produced by the percolation at the thresholds that produce the most relevant divisions (300m, 800m, 900m, 1400m, 5000m). 

If we were to suggest that the Scottish cluster coincided with predominant SNP voters, the south west with Liberal Democrats, the industrial north with Labour, and the south east  and shire counties with Conservatives, then one would not be far of the mark in what people have speculated this last six months about the forthcoming May election. UKIP do not show up in this physical decomposition and Wales blurs in the northern and western clusters for Plaid Cymru is largely a rural party, remote even within Wales. The local effects in urban areas where inner cities are more likely to vote Labour and the suburbs Conservative are picked up when we go down to the much finer city thresholds. It is this that encourages us that using percolation to detect the degree of isolationism as well as the strong concentration in the British population space of orientation and nearness to other areas are remarkably strong determinants of what people will vote. 

\section{Clustering of the percolation}
\subsection{Preliminary definitions}

Let $I$ be the set of intersections of the road network of Britain and let $C$ be a set of parliamentary constituencies with $n$ being the number of constituencies.
Each $C_i\subset C$ is a set of intersections that belong to the $i$-th constituency such that $\bigcup_{i=1}^n C_i=C$, $C_i\cap C_j=\emptyset$ for all $i$ and $j$ and $\bigcup_{i=1}^n\{x\mid x\in C_i\}=I$.

For each constituency $C_i$ we have a vector of voting behavior $\vec v_i=(v_{i,1},\cdots,v_{i,l})$ which is composed of $l$ elements, being $l$ the number of political parties. We will refer to the $k$-th element of the vector $\vec v_i$ as $v_{i,k}$, which is equal to the number of votes that the political party $k$ received in the constituency $i$.

\subsection{Obtaining the percolation clusters}
The technique to generate the percolation clusters is explained in detail in \cite{Arcaute2015}. We will explain in the following paragraphs a summarised version of the technique to perform a network percolation, how it serves to generate a tree of the percolations, and how we can use that tree to assign a unique cluster for each intersection of the road network of the UK.
Given a graph of the road network, where nodes represent intersections and the weight for each edge is the length of the street that connects them and a certain metric threshold (e.g. 5000m) we produce a network percolation by:
\begin{enumerate}
\item Selecting the transition of the graph with the smallest weight (distance), generating a new cluster and inserting both its nodes into the cluster.\item We will keep a first-in first-out queue of {\it nodes to expand}, from which we will extract a node to continue the process. We add both nodes of the transition selected in step 1 to this queue.
Nodes are only added to this queue if they are not already included.\item Extract a node from the queue of {\it nodes to explore} and if a transition departing from that node (not yet included in the cluster) is smaller than the threshold, include the transition in the cluster and the end node of the transition in the queue of nodes to explore.
\item Repeat step 3 until no further node can be expanded (the queue is empty) and if there are transitions left in the graph that do not belong to any cluster, generate a new cluster by choosing the smallest available transition and repeat from step 1. 
\end{enumerate}
This procedure will cover the complete graph with clusters, most of them irrelevant clusters of only a few nodes. To avoid this noisy behaviour we set a minimum size for a cluster of $75$ nodes in order to include it in the set of percolation clusters.

\begin{figure}[h]
\center{\includegraphics[width=1\linewidth]{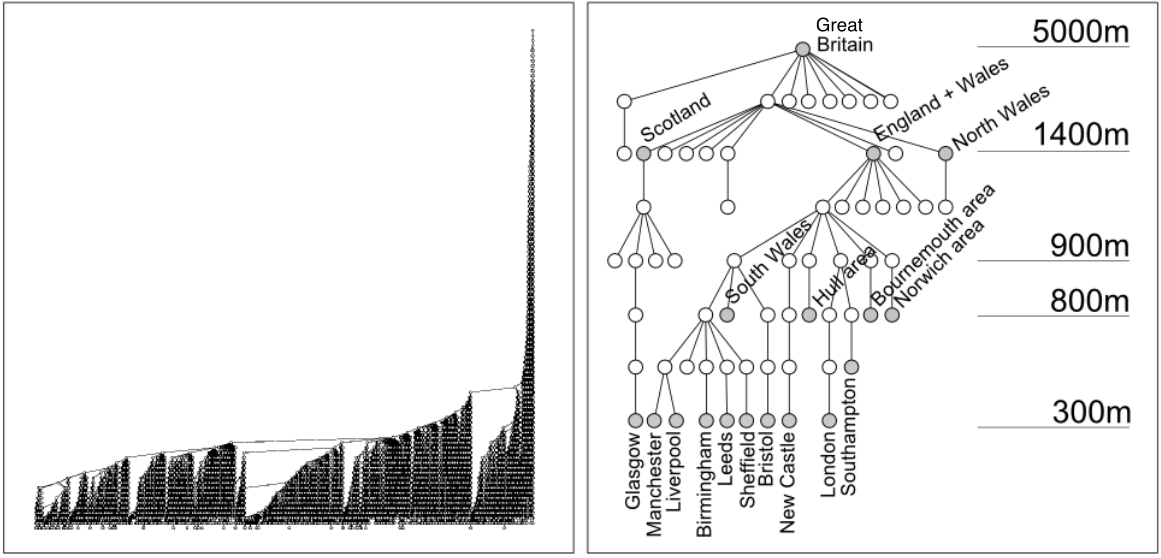}}
\caption{To the left, the complete tree of the percolation using all the calculated thresholds and every cluster larger than 75 nodes. To the right, a simplified version of the percolation tree, generated by using only some selected thresholds and the largest 10 clusters per distance presented to improve the understanding of the approach.}\label{treePercolation}
\end{figure}

If we repeat this algorithm for a large set of distance thresholds (in the interval $[5000,50]$ every $50$m), the largest distance will produce one single cluster for the whole of the UK that includes every intersection. The following distance will produce a set of smaller clusters completely contained in the previous one, leaving behind a few intersections. In fact, we can generate a tree of the percolation in this manner which renders the result portrayed in Figure~\ref{treePercolation}.

The set $P$ of percolation clusters is the extended set that includes every
cluster in this tree with $m$ being the number of percolation clusters. Each $P_j\subset P$ is a set of intersections that belong to the $j$-th percolation cluster such that $\bigcup_{j=1}^{m} P_j=P$ and $\bigcup_{j=1}^{m}\{x| x\in P_j\}=I$. The set $P$ is not a disjoint set, meaning that the same intersection will belong to several percolation clusters simultaneously as long as they have a parent-son relationship. That is, given two percolation clusters $P_j$ and $P_k$, $P_j\cap P_k\neq\emptyset$ if there exist a path in the percolation tree from $P_j$ to $P_k$ (or viceversa) and otherwise $P_j\cap P_k=\emptyset$. 

We will define the operation of intersecting a constituency with the set
of percolation clusters as the operation that returns a vector $(\vec p_i)$
with size $m$ where each component $j$ of the vector is the number of intersections
that belong to the intersection of $C_i\cap P_j$. That is, $C_i\cap P=\vec
p_i, p_{i,j}=\Vert \{x|x\in C_i\cap P_j\}\Vert $.

This operation serves to generate a vector for each constituency that determines its composition in terms of the percolation clusters and that in turn, will serve to cluster the constituencies into similar behavioral groups according to the percolation.\subsection{Clustering the percolation clusters following the parliamentary constituencies subdivision}

\begin{figure}[h]
\center{\includegraphics[width=1\linewidth]{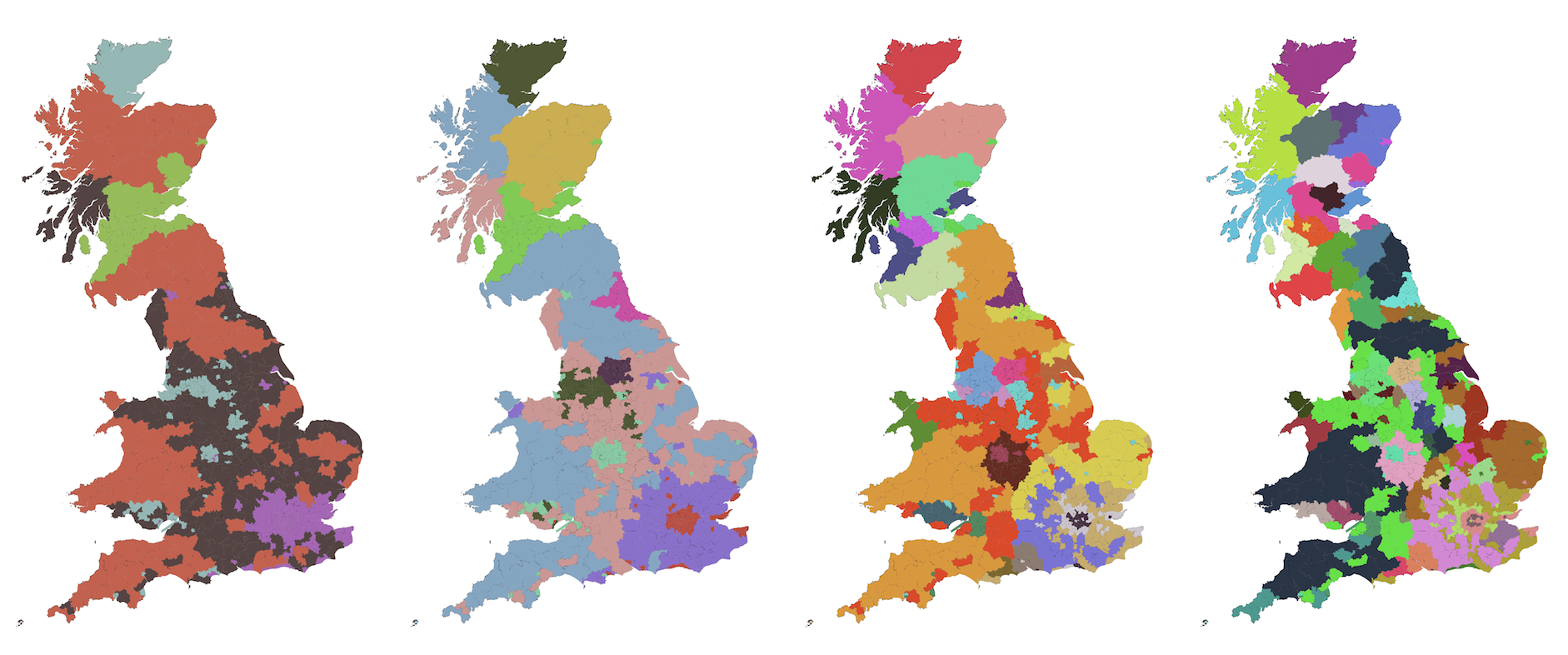}}
\caption{Partitioning around medoids of the constituencies in terms of their composition of network percolation clusters using $\{5,10,30,60\}$ clusters.}\label{60Clusters}
\end{figure}

We will use the partitioning around medoids algorithm (PAM \cite{Park2009}) to cluster the vectors $\vec p_i$ using the chi-squared distance \cite{PeleWerman2010} between them. In more detail, the distance between the constituencies $C_i$ and $C_k$ is $d(\vec p_i,\vec p_k)=\frac{1}{2}\sum_{\forall j}\frac{(p_{i,j}-p_{k,j})^2}{p_{i,j}+p_{k,j}}$. This algorithm will cluster the constituencies into different sets according to their composition of percolation clusters. We will call the set that holds this set of disjoint clusters $A$, where each component $A_g\subset A$ is composed of several constituencies ($A_g=\{C_i,C_k,\cdots\}$) such that $\bigcup_{g=1}^{\Vert A\Vert }A_g=C$ and $A_g\cap A_j=\emptyset$ for all $g$ and $j$. We can observe in Figure~\ref{60Clusters} the result of the applying the clustering algorithm to the space of the UK for different number of clusters. Throughout the rest of the paper we will use 60 clusters, given that it approximates about 10\% of the actual number of constituencies which represents a reasonable compression limit.

We can now extract a density of voting behaviour for each constituency that is given by the percolation by calculating the averaged voting behaviour of the densities of votes for each element of the set $A$. That is, given the $g$-th element of $A$, $$\vec v'_g=\frac{\sum_{i=1}^{\Vert A_g\Vert}\vec v_i'}{\Vert A_g\Vert }$$
where $\vec v_i'=\frac{\vec v_i}{\sum_{j=1}^l v_{i,j}}$ is the density vector of the voting behavior of the $i$-th constituency belonging to the set $A_g$.
\subsection{Performance of the clustering}

In order to quantify the quality of the partitioning of the constituencies set based on the percolation, we will study the error that it generates in comparison with several other types of partitioning.
On one side, we will take into account several socio-economic variables from the 2011 census \cite{CensusUK2011} that  could be relevant to the voting behaviour such as the degree of educational level (qualifications), the occupational class data (which serves as well as a proxy for income data), the age structure of the population and the country of birth (that can account for areas with a strong immigration rate).

The data of the different variables is normalized by the total number of people that each variable accounts for. Qualifications is a vector with 5 categories: no-qualification, level 1, 2, 3 and 4. Occupational data is a vector that distinguishes between  Managers, Professional, Associate Professional, Admin, Skilled Trades, Other Service, Sales, Process and Elementary. Age structure is another vector that has separated into components the number of people in each age group segregated by periods of 5 years and country of birth distinguishes between the categories UK, Ireland, Other EU, Other EU Accession and Rest of the World. We will later use the same partitioning mechanism to produce the clustering of the constituencies.

\begin{figure}[h]
\center{\includegraphics[width=1\linewidth]{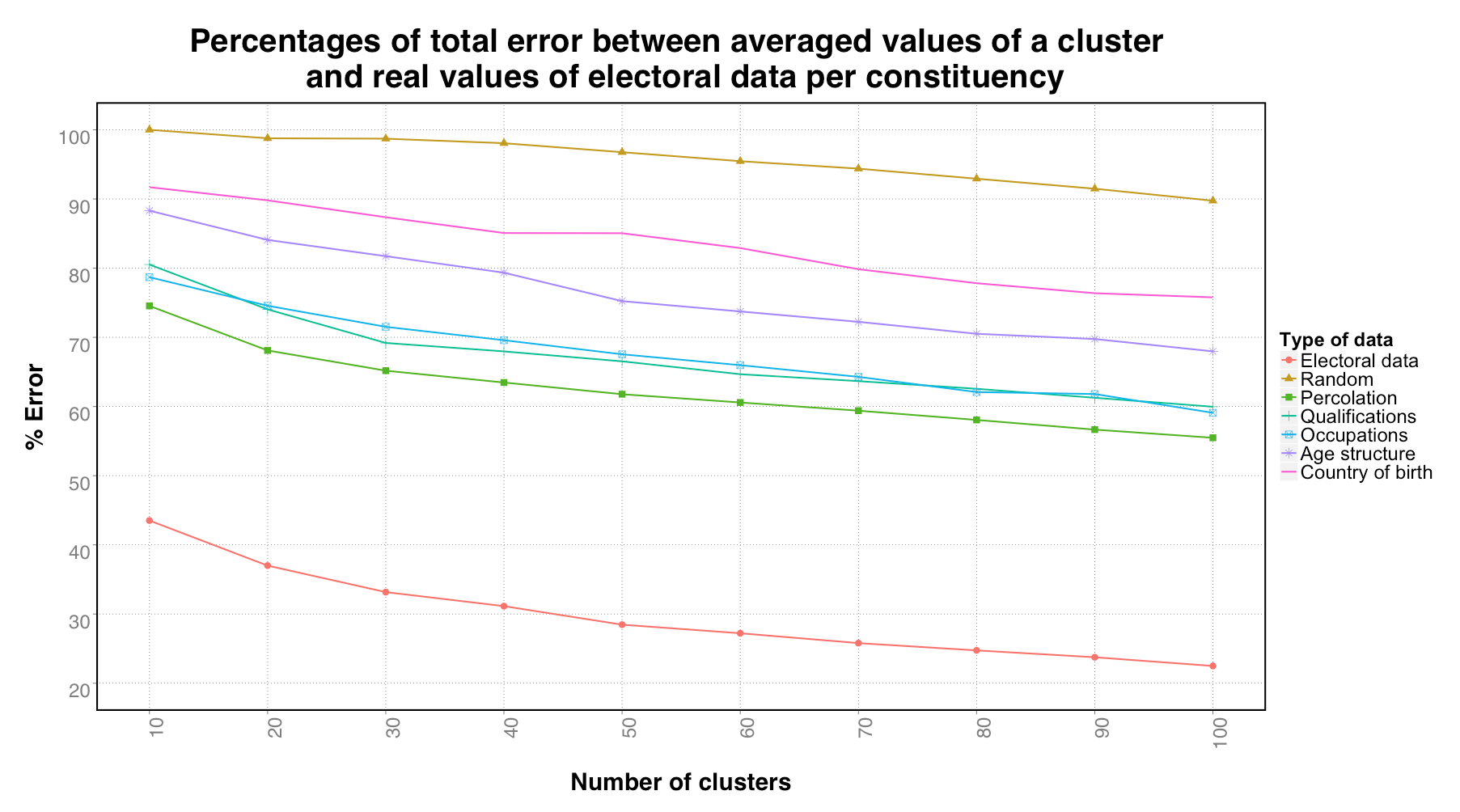}}
\caption{Plot of the percentage of total error for different number of clusters in which we are comparing the socio-economic data with the approach presented in this paper.\label{errors60clusters}}
\end{figure}

As we can observe in Figure~\ref{errors60clusters}, the best clustering corresponds to the percolation, with the second best being occupational class data and the qualification variable which performs similarly.
This shows how relevant it is the area to which one belongs, the level of connectivity that it has to other regions and how deep in the percolation tree a region is (cities have a larger depth than rural areas) to characterize voting behaviour. This can be explained relatively easy by considering the role that the inter-exchange of ideas between peers has on cultural patterns and that areas that are highly connected (or even belong to the same region) will have a wider range of migrant flows thus influencing each other's way of thinking. The full extent of this analysis and how to be able to improve the results by simultaneously using the socio-economic and the geographical data will be treated in future work.

 In order to produce the plot, we will calculate the error as the sum of the
distances between the averaged voting vector of $A_g$ ($\vec v_g'$) and the
voting behaviour of each constituency $C_i$ ($\vec v_i$) included in the
set $A_g$. That is:
$$error=\sum_{\forall g,\forall i} |w_{i}\cdot \vec v_g'-\vec v_i|$$
where $w_i=\sum_{\forall
j} v_{i,j}$ is the total number of votes for constituency $C_i$.

\begin{figure}[h]
\center{\includegraphics[width=1\linewidth]{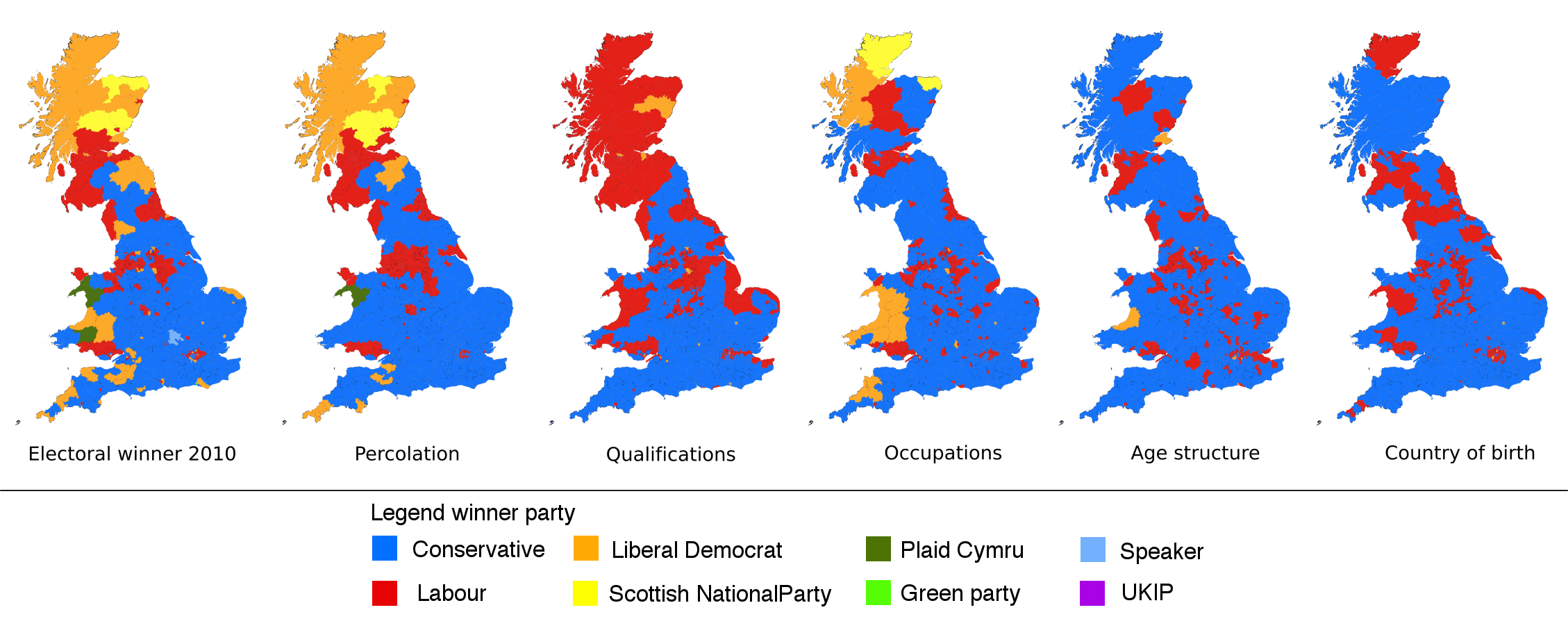}}
\caption{From left to right, actual results of the winners from the 2010 elections; winner extracted from the clustering based on the percolation; winner from the clustering of the qualifications variable; winner of the clustering from the occupational data; winners from age structure clustering and country of birth clustering.}\label{winner60Clusters}
\end{figure}

We can also extract the winner for each constituency using these averaged vectors to get an approximate idea of how well they separate the space in terms of voting behaviours as shown in Figure~\ref{winner60Clusters} for all the studied set of socio-economic data and the percolation methodology.

\subsection{Sub-trees}
We have produced the tree of the percolation using the full set of percolation thresholds from $5000$m to $50$m, but we could perform this procedure calculating the clusters from $5000$ to $4950$ and generating the tree for those thresholds, gradually decreasing the lowest threshold generating different sub-trees of increasing detail. We could then study their clusterings and their averaged voting behaviour and measure how much error each accounts for.

\begin{figure}[h]
\center{\includegraphics[width=1\linewidth]{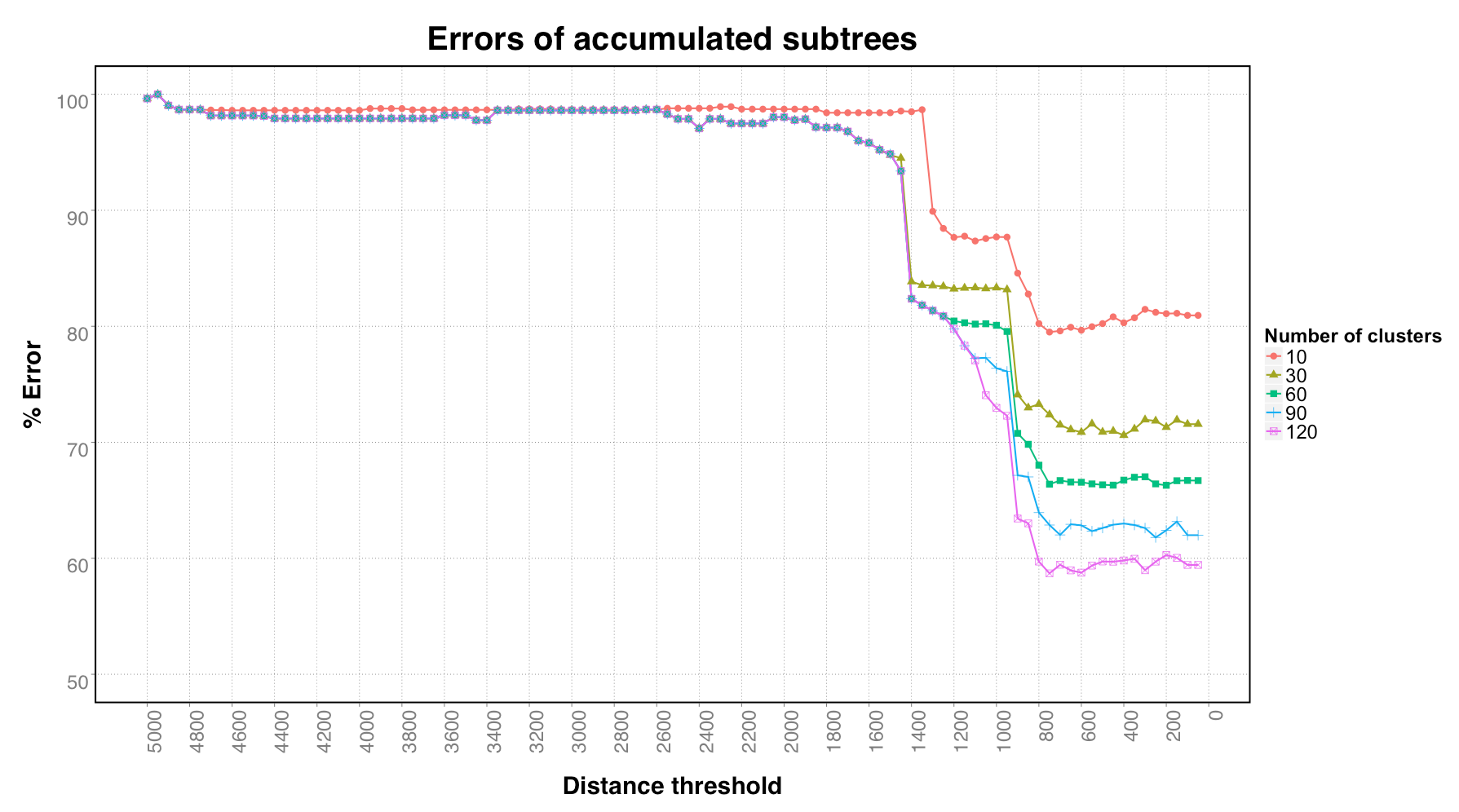}}
\caption{Errors of the accumulated subtrees for 60 clusters.}\label{errorsTrees}
\end{figure}

Furthermore, we can produce the plot presented in Figure~\ref{errorsTrees} for all the subtrees where the errors for the accumulated sub-trees are shown. In this Figure we can observe that there are 2 main thresholds where there are large decreases in the error thus producing local minima, exactly in the threshold of $1400$ and in the $900, 800$ range and in a smaller scale also in the $400, 300$ range. Those 3 thresholds correspond respectively to 3 scales, the nation scale, the regional scale and the city scale that were represented in Figure~\ref{percolations}.

\section{Predicting voting behaviours}
This entire approach is based on assuming that the percolation clusters identify a geographical pattern from which voter behaviour can emerge as a consequence of nationalistic and regional attitudes that reflect how Britain is fracturing into its long standing historical subdivisions. We should alert ourselves to the possibility that geographical factors are more of a determinant of the current volatility in voter attitudes than at any time in the last 100 years. To this end, we will examine how we might embed these geographical considerations into a simple model that is able to predict votes by combining the 2010 election results with our results from the percolation.

In order to predict voting behaviour we use the the uniform national swing method~\cite{Johnston1982} segregated by Scotland and England with Wales, which takes the following form:
\begin{enumerate}

\item Using the votes vectors of the constituencies ($\vec v_i$) we calculate the average votes for each party in the two areas (Scotland and England with Wales).
\item Taking into account the percentages published in the polls by {\it The New Statesman} of the final results (\url{http://may2015.com}) we produce a vector of swing votes from Scotland and England with Wales ($\vec s_i$ for the constituency $C_i$).
\item Using both vectors we can generate a new vector of predicted votes for each constituency $C_i$ as $\vec v_i^p=\vec v_i+\vec s_i$.
\end{enumerate}

In order to have control values for our methodology we use the predictions presented in  \url{http://www.electionforecast.co.uk/} which are shown in column $A_C$ of table~\ref{tableVotes} and to ensure that our simple methodology is capable of generating valuable results, we produce a prediction based on the actual votes ($\vec v_i$) from the 2010 elections which is shown in column $B_P$. As we can see in the table the results are quite similar.

 We then proceed to apply this method to generate a prediction based on the percolation. Instead of using the actual votes we use the averaged votes from the clusters generated with the percolation ($\vec v_i=w_i\cdot \vec v_g')$ and recalculate the swing votes to form column $D_P$. Finally, we do the same for the clusters generated from the occupational data to generate column $E_P$ and later on, we calculate the average between the votes obtained with the percolation and the votes obtained with the occupational data,  substitute $v_i$ to recalculate the swing votes and produce the output shown in column $C_P$.

As we can observe, the result obtained with the percolation clusters overestimate the impact of the Labour Party while the occupational data produces the inverse effect. Using the average of both voting behaviours produces a map of constituencies that resemble to a large extent the actual voting prediction showing that there is a underlying relationship between the voting patterns of people and their location in the network in close relationship with the occupational data. 

\begin{figure}[h]
\center{\includegraphics[width=1\linewidth]{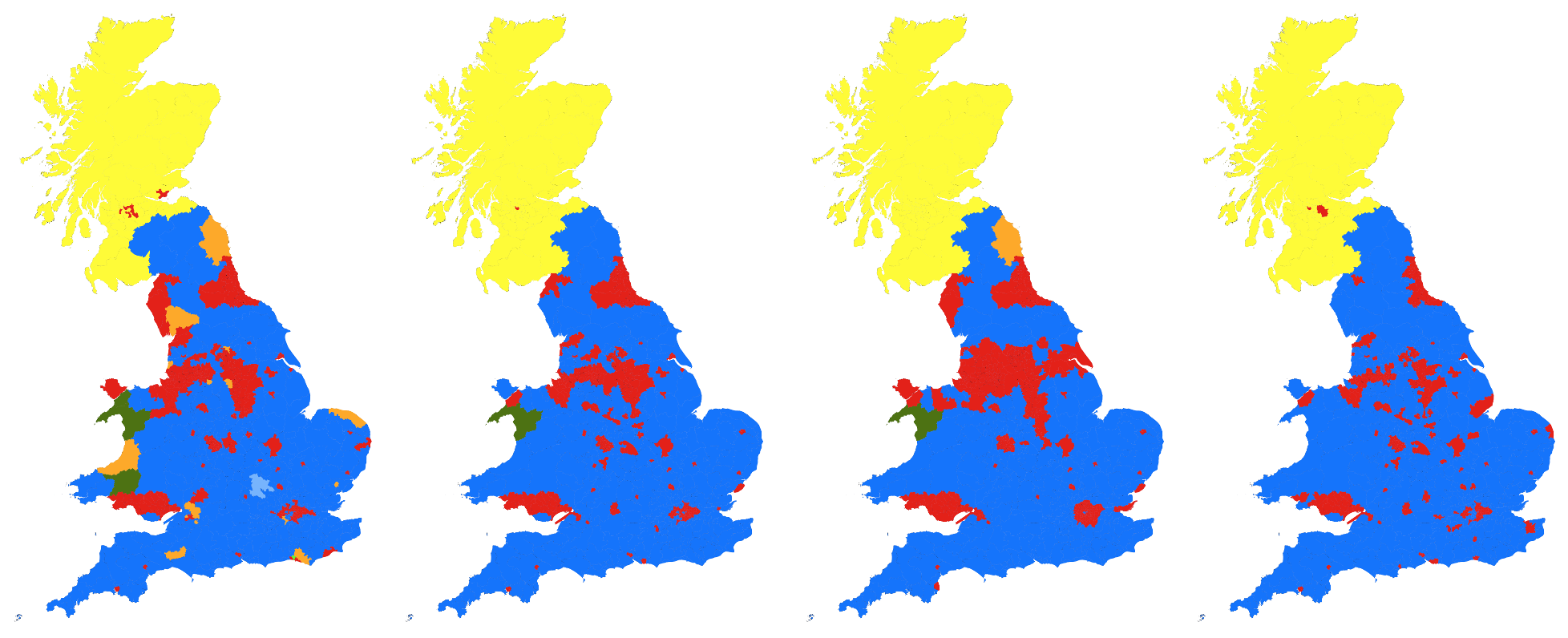}}
\caption{From left to right: ($B_P$) Prediction using the actual votes and the polls; ($C_P$) prediction by using the percolation and the occupational data; ($D_P$) prediction using solely the percolation; and ($E_P$) prediction using the occupational data.}\label{votingPatterns}

\end{figure}

\begin{table}
\begin{center}
\scalebox{.7}{

\begin{tabular}{ |l|l|l|l|l|l| }

  \hline
  Parties&$A_C$&$B_P$&$C_P$&$D_P$&$E_P$\\
  
  \hline
  
    Conservatives & 286 &283&301&260&323\\
  Labour & 267&276&272&311&252 \\
  SNP & 48&49&55&56&54 \\
  Liberal Democrats & 24 &16&0&1&0\\
  Plaid Cymru & 4 &3&1&1&0\\ 
  Greens &1&1&0&0&0\\
  UKIP&1&0&0&0&0\\
 Speaker & 1 &1&0&0&0\\\hline

\end{tabular} 

}
\end{center}
\caption{Voting predictions by number of seats. $A_C$ prediction from \url{http://www.electionforecast.co.uk/}. $B_P$ Prediction based on the real voting vectors. $C_P$ prediction based on the percolation and the occupational class data voting behaviour. $D_P$ prediction based on the percolation voting behaviour. $E_P$ prediction based in the occupational data. \label{tableVotes}}
\end{table}

\section{So What Will be the Outcome?}

In essence, our model does not pick up either the extremes of voting that the current polls are showing, nor does it produce the neck and neck race between the traditional parties. If the result of the election is as the recent YouGov polls suggest (see \url{http://www.electionforecast.co.uk/})  the Conservatives will gain 285 seats, Labour around 270, the SNP 50 and the Liberal Democrats 25. This indeed would be a strange result by historical standards. It probably represents a hung parliament with no party able to win an outright majority and in fact no parties able to form a stable coalition. The closest that our model comes to forecasting this is with the Conservatives on 301, Labour 271, the SNP 56 while the Liberal Democrats are erased from the map. But our most extreme prediction which still takes account of the geographical effects produces a much larger concentration of seats for the Labour Party which brings up the topic of how capable is the system of first pass the post to represent proportionally the number of votes taking into account that different partitioning of the constituencies produce very different results. This variation is present as well in the opposite range by the partition generated from the occupational data which agglomerate the constituencies in such a way that the Conservatives get a clear win. Strange times indeed. 

To an extent what we have developed here is a work in progress. We will only be able to refine our model, once the votes are known on May 7th 2015, when we will be able to undertake a much more considered analysis of geographical factors but we remain convinced that geographical isolation, separation, and connectivity is a key factor in determining not only how people vote but even how they think and it is this that would appear to be dictating the high volatility of current  and more considered predictions. In fact in such a situation, there could well be a final bounce or shift, a transition to traditional or even more extreme or some combination of both when the voters take to polls and the votes are finally counted. 

Exciting times.

\newpage

\bibliographystyle{elsarticle-num}

\begin{thebibliography}{10}
\expandafter\ifx\csname url\endcsname\relax
  \def\url#1{\texttt{#1}}\fi
\expandafter\ifx\csname urlprefix\endcsname\relax\def\urlprefix{URL }\fi
\expandafter\ifx\csname href\endcsname\relax
  \def\href#1#2{#2} \def\path#1{#1}\fi

\bibitem{Arcaute2015}
E.~Arcaute, C.~Molinero, E.~Hatna, R.~Murcio, C.~Vargas-Ruiz, P.~Masucci,
  J.~Wang, M.~Batty, {Hierarchical organisation of Britain through percolation
  theory} (2015) 11, \href {http://arxiv.org/abs/1504.08318}
  {\path{arXiv:1504.08318}}.


\bibitem{Johnston2006}
R.~Johnston, C.~Pattie, {Putting voters in their place: Geography and elections
  in Great Britain}, Oxford University Press, 2006.

\bibitem{Johnston2009}
R.~Johnston, C.~Pattie, {Geography: The Key to Recent British Elections},
  Geography Compass~(5)  1865--1880.
\newblock \href {http://dx.doi.org/10.1111/j.1749-8198.2009.00258.x}
  {\path{doi:10.1111/j.1749-8198.2009.00258.x}}.

\bibitem{Perez2015}
T.~P\'erez, J.~Fern\'andez-Gracia, J.~J. Ramasco, V.~M. Egu\'iluz, Persistence
  in voting behavior: Stronghold dynamics in elections, in: N.~Agarwal, K.~Xu,
  N.~Osgood (Eds.), Social Computing, Behavioral-Cultural Modeling, and
  Prediction, Vol. 9021 of Lecture Notes in Computer Science, Springer
  International Publishing, 2015, pp. 173--181.
\newblock \href {http://dx.doi.org/10.1007/978-3-319-16268-3{\_}18}
  {\path{doi:10.1007/978-3-319-16268-3{\_}18}}.

\bibitem{Torok2013}
J.~T\"{o}r\"{o}k, G.~I\~{n}iguez, T.~Yasseri, M.~{San Miguel}, K.~Kaski,
  J.~Kert\'{e}sz, {Opinions, Conflicts, and Consensus: Modeling Social Dynamics
  in a Collaborative Environment}, Physical Review Letters 110~(8) (2013)
  088701.
\newblock \href {http://dx.doi.org/10.1103/PhysRevLett.110.088701}
  {\path{doi:10.1103/PhysRevLett.110.088701}}.

\bibitem{Stauffer2002}
D.~Stauffer, {Percolation and Galam Theory of Minority Opinion Spreading},
  International Journal of Modern Physics C 13~(07) (2002) 975--977.
\newblock \href {http://dx.doi.org/10.1142/S0129183102003735}
  {\path{doi:10.1142/S0129183102003735}}.

\bibitem{Shao2009}
J.~Shao, S.~Havlin, H.~E. Stanley, {Dynamic Opinion Model and Invasion
  Percolation}, Physical Review Letters 103~(1) (2009) 018701.
\newblock \href {http://dx.doi.org/10.1103/PhysRevLett.103.018701}
  {\path{doi:10.1103/PhysRevLett.103.018701}}.

\bibitem{Balankin2011}
A.~S. Balankin, M.~A. {Mart\'{\i}nez Cruz}, A.~T. Mart\'{\i}nez, {Effect of
  initial concentration and spatial heterogeneity of active agent distribution
  on opinion dynamics}, Physica A: Statistical Mechanics and its Applications
  390~(21-22) (2011) 3876--3887.
\newblock \href {http://dx.doi.org/10.1016/j.physa.2011.05.034}
  {\path{doi:10.1016/j.physa.2011.05.034}}.

\bibitem{Travieso2006}
G.~Travieso, L.~{da Fontoura Costa}, {Spread of opinions and proportional
  voting}, Physical Review E 74~(3) (2006) 036112.
\newblock \href {http://arxiv.org/abs/0603044} {\path{arXiv:0603044}}, \href
  {http://dx.doi.org/10.1103/PhysRevE.74.036112}
  {\path{doi:10.1103/PhysRevE.74.036112}}.

\bibitem{Lambiotte2007}
R.~Lambiotte, M.~Ausloos, J.~A. Hołyst, {Majority model on a network with
  communities}, Physical Review E 75~(3) (2007) 030101.
\newblock \href {http://dx.doi.org/10.1103/PhysRevE.75.030101}
  {\path{doi:10.1103/PhysRevE.75.030101}}.

\bibitem{Park2009}
H.-S. Park, C.-H. Jun, {A simple and fast algorithm for K-medoids clustering},
  Expert Systems with Applications~(2)  3336--3341.
\newblock \href {http://dx.doi.org/10.1016/j.eswa.2008.01.039}
  {\path{doi:10.1016/j.eswa.2008.01.039}}.

\bibitem{PeleWerman2010}
The quadratic-chi histogram distance family, in: K.~Daniilidis, P.~Maragos,
  N.~Paragios (Eds.), Computer Vision – ECCV 2010, Vol. 6312 of Lecture Notes
  in Computer Science, 2010.
\newblock \href {http://dx.doi.org/10.1007/978-3-642-15552-9_54}
  {\path{doi:10.1007/978-3-642-15552-9_54}}.

\bibitem{CensusUK2011}
\href{https://www.nomisweb.co.uk/census/2011}{Office for national
  statistics. 2011 census} (2011).
\newline\urlprefix\url{https://www.nomisweb.co.uk/census/2011}

\bibitem{Johnston1982}
R.~J. Johnston, A.~M. Hay, {On the parameters of uniform swing in single-member
  constituency electoral systems}, Environment and Planning A 14~(1) (1982)
  61--74.
\newblock \href {http://dx.doi.org/10.1068/a140061}
  {\path{doi:10.1068/a140061}}.

\end{thebibliography}

\end{document}